\documentclass{article}
\twocolumn

\usepackage{graphicx}
\usepackage{amsthm}
\usepackage{amsmath,amssymb}
\usepackage[utf8]{inputenc}
\usepackage[T2A]{fontenc}

\AtBeginDocument{\addtolength{\columnsep}{8pt}}
\AtBeginDocument{\addtolength{\textwidth}{10pt}}
\begin{document}

\title{\vspace{1cm}{\Large {\bf Multistrand Eigenvalue conjecture and Racah symmetries  }\vspace{.2cm}}}
% \author{\textbf{Liudmila Bishler}\footnote{ {\small \textit{LPI} and \textit{ITEP, Moscow, Russia}};
% mila-bishler@mail.ru}, \ \textbf{Andrey Morozov}\thanks{{\small \textit{Moscow State University} and \textit{ITEP, Moscow, Russia}};
% Andrey.Morozov@itep.ru}\date{ }}
	\author{{\bf Andrey Morozov$^{a,b}$\thanks{e-mail: morozov.andrey.a@iitp.ru}}
\\
	$^a$ {\small {\it NRC Kurchatov Institute}}\\
	$^b$ {\small {\it Institute for Information Transmission Problems}}
\\
\vspace{-5.5cm}
\\
\hfill\hspace{13cm}ITEP/TH-26/22
\\
\hfill IITP/TH-23/22
\vspace{5cm}
}
\date{
\begin{minipage}[b]{31pc}  
%\begin{raggedleft}
\begin{abstract}
Racah matrices of quantum algebras are of great interest at present time. These matrices have a relation with $\mathcal{R}$-matrices, which are much simpler than the Racah matrices themselves. This relation is known as the eigenvalue conjecture. In this paper we study symmetries of Racah matrices which follow from the eigenvalue conjecture for multistrand braids.
\end{abstract}
%\end{raggedleft}
\end{minipage}
}

\maketitle

\section{Introduction}

Racah matrices and $6j$-symbols have quite a long history in science. They first appeared in physics through the studies of angular momenta \cite{R1}, where they describe transformation between different couplings of three particles. Racah matrices appearing in this problem are actually quite simple and well studied. However, in mathematical terms this angular momenta problem actually corresponds to the multiplication of different representations of the $SU(2)$ algebra. In this language Racah coefficients appear when one takes a product of three representations and expands it as a sum of irreducible representations:
\begin{equation}
T_1\otimes T_2 \otimes T_3=\sum M_{12}^{1,2} Q_{12} \otimes T_3= \sum N^{12,3}M_{12}^{1,2} Q.
\end{equation}
There is also another way to represent the same product:
\begin{equation}
T_1\otimes T_2 \otimes T_3=\sum T_1 \otimes M_{23}^{2,3} Q_{23}= \sum N^{1,23}M_{23}^{2,3} Q.
\end{equation}
Map betwenn these two bases is called Racah matrix, which elements are $6j$-symbols:
\begin{equation}
\begin{array}{c}
U:\ N^{12,3}M_{12}^{1,2}\rightarrow N^{1,23}M_{23}^{2,3},
\\
\\
U^{12}_{23}=\left\{
\begin{array}{ccc}
T_1 & T_2 & Q_{12}
\\
T_3 & Q & Q_{23}
\end{array}
\right\}
\end{array}
\end{equation}

What is interesting about such a definition is that it can be easily generalized to the representations of other algebras, which makes them applicable to other physical theories such as chromodynamics, where $SU(3)$ group should be used. In this case representations should be enumerated not by natural numbers as in $SU(2)$ case, but by Young diagrams $T=[T_1,\ldots T_n]$, $T_1\geq T_2 \geq T_3 \ldots$. However, one can generalize this definition to not only other Lie algebras, but also to their deformations such as quantum groups \cite{Klimyk}. Quantum groups are essentially a deformation of universal enveloping algebras. They are connected among other things  integrable systems \cite{R3}, lattice gauge theory \cite{R6}, 3-d quantum gravity \cite{R7}, quantum $sl_N$ invariants of knots \cite{R8}, Turaev-Viro invariants of 3-manifolds and topological field theory \cite{R9,R10}, WZW conformal field theory and 3d Chern-Simons theory \cite{R4,R5}. In the latter case quantum groups describe the hidden symmetries of the observables, which are also equal to the knot polynomials from the knot theory.

\begin{figure}[h!]
\begin{center}
\includegraphics[width=0.45\textwidth]{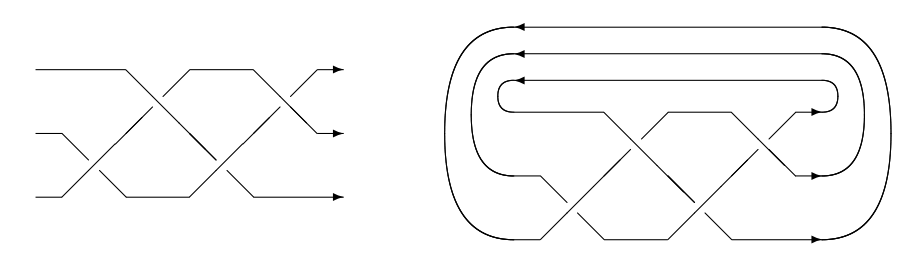}
\end{center}
\caption{Braid representation of a knot $4_1$, braid is on the left while on the right is it closure which is a knot.
\label{f:braid}}
\end{figure}

In recent years we actively studied applications of quantum groups and Racah matrices to the calculations of knot polynomials and observables in Chern-Simons theory. The main problem of these approaches is that one has to know many different Racah matrices to calculate corresponding knot polynomials. However, at the moment there are only few cases where general answers for Racah matrices are known. Among these are Racah matrices for $U_q(sl_2)$ group \cite{sl2} and multiplicity free exclusive Racah matrices \cite{NonMult1,NonMult2}. In the context of the present paper we are interested in Racah matrices which appear in the calculations using braid representation of a knot, see Fig.\ref{f:braid}.

In \cite{Eig} quite a powerful tool to find missing Racah matrices was suggested, namely the eigenvalue conjecture which relates known eigenvalues of quantum $\mathcal{R}$-matrices with the Racah matrices for the 3-strand braids. This conjecture was further studied in \cite{EigAl,Cab,EigSym,SymProof}. Since eigenvalue conjecture relates different Racah matrices, it imposes certain symmetries on Racah matrices, which were studied in \cite{EigSym,SymProof}. However, symmetries studied in these papers are the symmetries connected with the three-strand eigenvalue conjecture. There is also eigenvalue conjecture for higher number of strands \cite{MultEig}. In the present letter we are interested in studying symmetries implied by this multi-strand eigenvalue conjecture.

The paper is organized as follows. In Section 2 we briefly remind the structure of Reshetikhin-Turaev approach to the knot calculus, involving $\mathcal{R}$ and Racah matrices. In Section 3 we explain why Racah matrices for multistrand braids possess certain block structure. In Section 4 we remind the formulation of eigenvalue conjecture. In Section 5 the main result of the paper, (\ref{e:4eig}) and (\ref{e:5eig}), is introduced and explained.

\section{Reshetikhin-Turaev approach}

One of the most productive approaches to the calculations of knot polynomials (as well as Wilson-loop averages of the Chern-Simons theory, which are equal to them) is modern Reshetikhin-Turaev approach \cite{RT}-\cite{RTfin}. It is most clearly defined for the braid representation of a knot (see Fig.\ref{f:braid}). Then in such a braid all crossings correspond to $\mathcal{R}$-matrices. These $\mathcal{R}$-matrices are very simple in their diagonal form. Eigenvalues of the $\mathcal{R}$-matrices correspond to irreducible representations in a tensor product of the representations corresponding to two crossing strands. For the knot case and representation $Q$ from the product of two representations $T$ the eigenvalue is equal to \cite{Reig1,Reig2,Reig3,Klimyk}
\begin{equation}
\label{e:eig}
\lambda_Q=q^{\varkappa_Q-4\varkappa_T-N|T|},
\end{equation}
where $\varkappa_Q$ is calculated directly from the Young diagram $Q$:
\begin{equation}
\varkappa_Q=\sum\limits_{\{i,j\}\in Q}(i-j),
\end{equation}
where $\{i,j\}$ enumerates the boxes in the Young diagram.

For two-strand braids, which are parameterised by a number of crossings $n$, the answer for the knot polynomial is then given by a character expansion:

\begin{equation}
H^{T[2,n]}_T=\sum\limits_{Q\vdash T\otimes T} S^*_Q(A,q) Tr \mathcal{R}^n_Q=\sum\limits_{Q\vdash T\otimes T} S^*_Q(A,q) \lambda^n_Q,
\end{equation}
where $S^*_Q(A,q)$ are Shur polynomials (characters of the representation $Q$ for the quantum group) in a special point \cite{RTfin}, which can also be easily calculated from the Young diagram.

For multi-strand braids situation is more complicated. There are $\mathcal{R}$-matrices, corresponding to the crossings of different pairs of strands. They still can be made diagonal and thus provided by the (\ref{e:eig}), but then rotation matrices which change basis are needed. These rotation matrices are Racah matrices. Then answer for polynomials of $m$ strand braids are provided by the modified version of character expansion:
\begin{equation}
H^{\mathcal{K}}_T=\sum\limits_{Q\vdash T^{\otimes m}} S^*_Q(A,q) Tr \mathcal{B}_Q,
\end{equation}
where $\mathcal{B}_Q$ is a product of diagonal $\mathcal{R}$-matrices and Racah matrices, which order can be written down from the braid. This means that the most difficult quantities to find in this whole approach are Racah matrices.

\section{Block structure}

Racah matrices and, correspondingly, non-diagonal $\mathcal{R}$-matrices for the number of strands larger than 3 possess a certain block structure \cite{Cab,MultEig}.

\begin{figure}[h!]
\begin{picture}(200,75)(-55,-25)
\put(0,0){\line(-1,1){30}}
\put(0,0){\line(1,1){30}}
\put(-15,15){\line(1,1){15}}
\put(-33,32){\mbox{$T$}}
\put(-3,32){\mbox{$T$}}
\put(27,32){\mbox{$T$}}
\put(-20,0){\mbox{$Q_{12}$}}
\put(-5,-15){\mbox{$Q$}}
\put(0,0){\line(1,-1){15}}
\put(45,0){\mbox{$= \sum_{Q_{23}} u^{Q_{12}}_{Q_{23}} $}}
\put(130,0){
\put(0,0){\line(-1,1){30}}
\put(0,0){\line(1,1){30}}
\put(15,15){\line(-1,1){15}}
\put(-33,32){\mbox{$T$}}
\put(-3,32){\mbox{$T$}}
\put(27,32){\mbox{$T$}}
\put(10,0){\mbox{$Q_{23}$}}
\put(-5,-15){\mbox{$Q$}}
\put(0,0){\line(1,-1){15}}
}
\end{picture}
\caption{Pictorial description of 3-strand Racah matrix.\label{f:3Rac}}
\end{figure}
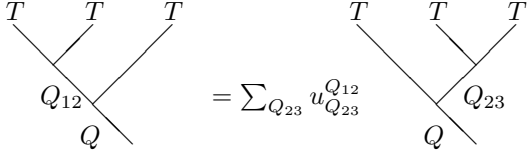

For 3-strand braid if one wants to contruct non-diagonal $\mathcal{R}$-matrix for the construction of $\mathcal{B}_Q$, the following Racah matrix is required:
\begin{equation}
\mathcal{R}_2=U\mathcal{R}U^{\dagger},\ \ \ U=\left[
\begin{array}{ccc}
T & T & Q_{12} \\
T & Q & Q_{23}
\end{array}\right].
\label{e:3Rac}
\end{equation}
This Racah matrix can also be described pictorially, see Fig. \ref{f:3Rac}.

\begin{figure}[h!]
\begin{picture}(200,85)(-40,-45)
\put(0,0){\line(-1,1){30}}
\put(0,0){\line(1,1){30}}
\put(-15,15){\line(1,1){15}}
\put(-33,32){\mbox{$T$}}
\put(-3,32){\mbox{$T$}}
\put(27,32){\mbox{$T$}}
\put(-20,0){\mbox{$Q_{12}$}}
\put(0,0){\line(1,-1){15}}
\put(-10,-15){\mbox{$Q_{123}$}}
\put(15,-15){\line(1,1){45}}
\put(57,32){\mbox{$T$}}
\put(15,-15){\line(1,-1){15}}
\put(27,-40){\mbox{Q}}
\put(45,0){\mbox{$= \sum_{Q_{23}} u^{Q_{12}}_{Q_{23}} $}}
\put(130,0){
\put(0,0){\line(-1,1){30}}
\put(0,0){\line(1,1){30}}
\put(15,15){\line(-1,1){15}}
\put(-33,32){\mbox{$T$}}
\put(-3,32){\mbox{$T$}}
\put(27,32){\mbox{$T$}}
\put(10,0){\mbox{$Q_{23}$}}
\put(0,0){\line(1,-1){15}}
\put(-10,-15){\mbox{$Q_{123}$}}
\put(15,-15){\line(1,1){45}}
\put(57,32){\mbox{$T$}}
\put(15,-15){\line(1,-1){15}}
\put(27,-40){\mbox{Q}}
}
\end{picture}
\caption{Pictorial description of 4-strand Racah matrix $U$.
\label{f:4Rac1}}
\end{figure}
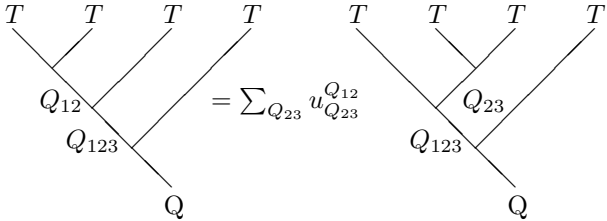

When we move to the 4-strands, two distinct types of Racah matrices are needed. First, to find $\mathcal{R}_2$, according to Fig. \ref{f:4Rac1}, similar matrix $U_4$ is needed. One should write irreducible representations expansion for the product of four representations $T$ again in two ways:
\begin{equation}
\begin{array}{l}
\left((T\otimes T)\otimes T\right)\otimes T=\!\!\!\sum\limits_{Q_{12}\in T\otimes T}\!\!\! \left(Q_{12}\otimes T\right)\otimes T=
\\ \\
=\sum\limits_{Q_{12}\in T\otimes T}\sum\limits_{Q_{123}\in Q_{12}\otimes T} Q_{123}\otimes T=\sum\limits_{Q\in T^{\otimes 4}}Q,
\end{array}
\end{equation}
and
\begin{equation}
\begin{array}{l}
\left(T\otimes (T\otimes T)\right)\otimes T=\!\!\!\sum\limits_{Q_{23}\in T\otimes T}\!\!\! \left(T\otimes Q_{23}\right)\otimes T=
\\ \\
=\sum\limits_{Q_{23}\in T\otimes T}\sum\limits_{Q_{123}\in T\otimes Q_{23}} Q_{123}\otimes T=\sum\limits_{Q\in T^{\otimes 4}}Q.
\end{array}
\label{e:b2}
\end{equation}
Racah matrix $U_4$ which describes rotation between these two bases consists of blocks, since it does not mix representations, which come from different $Q_{123}$. This means that in fact these blocks are in fact Racah matrices $U$ from (\ref{e:3Rac}), with $Q$ substituted by $Q_{123}$. Then $\mathcal{R}_2=U_4\mathcal{R}U^{\dagger}_4$.

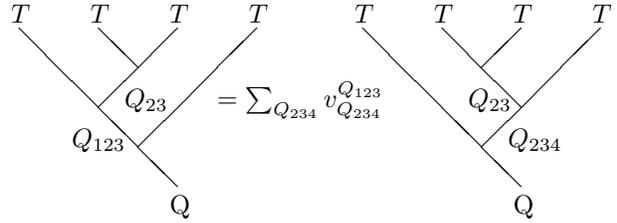
\begin{figure}[h!]
\begin{picture}(200,85)(-40,-40)
\put(0,0){\line(-1,1){30}}
\put(0,0){\line(1,1){30}}
\put(15,15){\line(-1,1){15}}
\put(-33,32){\mbox{$T$}}
\put(-3,32){\mbox{$T$}}
\put(27,32){\mbox{$T$}}
\put(10,0){\mbox{$Q_{23}$}}
\put(0,0){\line(1,-1){15}}
\put(-10,-15){\mbox{$Q_{123}$}}
\put(15,-15){\line(1,1){45}}
\put(57,32){\mbox{$T$}}
\put(15,-15){\line(1,-1){15}}
\put(27,-40){\mbox{Q}}%
\put(45,0){\mbox{$= \sum_{Q_{234}} v^{Q_{123}}_{Q_{234}} $}}
\put(130,0){
\put(0,0){\line(-1,1){30}}
\put(30,0){\line(-1,1){30}}
\put(15,15){\line(1,1){15}}
\put(-33,32){\mbox{$T$}}
\put(-3,32){\mbox{$T$}}
\put(27,32){\mbox{$T$}}
\put(10,0){\mbox{$Q_{23}$}}
\put(0,0){\line(1,-1){15}}
\put(25,-15){\mbox{$Q_{234}$}}
\put(15,-15){\line(1,1){45}}
\put(57,32){\mbox{$T$}}
\put(15,-15){\line(1,-1){15}}
\put(27,-40){\mbox{Q}}
}
\end{picture}
\caption{Pictorial description of 4-strand Racah matrix $V$.
\label{f:4Rac2}}
\end{figure}

To find $\mathcal{R}_3$ additional Racah matrix $V$ is required. The natural way is to look at the matrix corresponding to the Fig.\ref{f:4Rac2}. In the language of irreducible representation expansion this matrix connects bases (\ref{e:b2}) and
\begin{equation}
\begin{array}{l}
T\otimes \left((T\otimes T)\otimes T\right)=\!\!\!\sum\limits_{Q_{23}\in T\otimes T}\!\!\! T\otimes \left(Q_{23}\otimes T\right)=
\\ \\
=\sum\limits_{Q_{23}\in T\otimes T}\sum\limits_{Q_{234}\in T\otimes Q_{23}} T\otimes Q_{234}=\sum\limits_{Q\in T^{\otimes 4}}Q.
\end{array}
\label{e:b3}
\end{equation}
This matrix again consists of blocks. These blocks do not mix different representations $Q_{23}$ and correspond to following blocks of $6j$-symbols
\begin{equation}
V_i=\left[
\begin{array}{ccc}
T & Q_{23} & Q_{123} \\
T & Q & Q_{234}
\end{array}\right].
\end{equation}
The resulting $\mathcal{R}$-matrix is then equal to
\begin{equation}
\mathcal{R}_4=U_4VU_4\mathcal{R}\left(U_4VU_4\right)^{\dagger}.
\end{equation}
Another interesting thing about matrix $V$ is that since it does not mix different representations $Q_{23}$, and eigenvalues of $\mathcal{R}$-matrix depend on these representations, they in fact commute $\mathcal{R}V=V\mathcal{R}$. This structure in fact continues further to larger number of strands with each new strand adding new Racah matrix.

\section{Eigenvalue conjecture}

The origins if the eigenvalue conjecture comes from studying Yang-Baxter equation on $\mathcal{R}$-matrices:
\begin{equation}
\mathcal{R}_1\mathcal{R}_2\mathcal{R}_1=\mathcal{R}_2\mathcal{R}_1\mathcal{R}_2.
\end{equation}
If one rewrites it as an equation on diagonal $\mathcal{R}$-matrix and Racah matrices, then it looks like this:
\begin{equation}
\mathcal{R}U\mathcal{R}U^{\dagger}\mathcal{R}=U\mathcal{R}U^{\dagger}\mathcal{R}U\mathcal{R}U^{\dagger}.
\label{YBU}
\end{equation}
This equation can be interpreted as an equation on Racah matrices $U$. This leads to the eigenvalue conjecture \cite{Eig}:

\textbf{If normalied eigenvalues of two $\mathcal{R}$-matrices coincide then the corresponding Racah matrices also coincide.}

We are interested only in normalized eigenvalues, since if all $\mathcal{R}$-matrices in (\ref{YBU}) are multiplied by some constant the equation still holds.

However (\ref{YBU}) includes only three-strand Racah matrices. There are also other Racah matrices which appear for larger number of strands. As it appears \cite{MultEig}, other Yang-Baxter equations such as $\mathcal{R}_3\mathcal{R}_2\mathcal{R}_3=\mathcal{R}_2\mathcal{R}_3\mathcal{R}_2$ are actually trivially satisfied if the first Yang-Baxter equation is satisfied. The important equations in the multi-strand case are commutation relations for $\mathcal{R}$-matrices, such as $\mathcal{R}_1\mathcal{R}_3=\mathcal{R}_3\mathcal{R}_1$, or in the terms of diagonal $\mathcal{R}$-matrices
\begin{equation}
\mathcal{R}UV_1U\mathcal{R}(UV_1U)^{\dagger}=UV_1U\mathcal{R}(UV_1U)^{\dagger}\mathcal{R}.
\end{equation}
Also it is important to note that these Racah matrices possess certain block structure, as was described in the previous section. This leads to the following generalization of the eigenvalue conjecture \cite{MultEig}:

\textbf{If normalized eigenvalues of two $\mathcal{R}$-matrices coincide and the block structure of the corresponding multistrand Racah matrices coincide then the Racah matrices themselves also coincide.}

\section{Eigenvalue conjecture and symmetries of Racah matrices}

For 3-strand braids in symmetric representations it was first observed \cite{Eig} and later proven \cite{EigSym,SymProof}, that from the Eigenvalue conjecture it follows that following Racah matrices coincide:
\begin{equation}
\label{e:3eig}
%\begin{array}{l}
\left[
\begin{array}{ccc}
\ [r] & [r] & [a,b] \\
\ [r] & [l,m,n] & [x,y]
\end{array}\right]
=%\\
\end{equation}
%\\
\begin{equation*}\left[
\begin{array}{ccc}
\ [r+c] & [r+c] & [a+c,b+c] \\
\ [r+c] & [l+c,m+c,n+c] & [x+c,y+c]
\end{array}\right].
%\end{array}
\end{equation*}
In fact no other Racah matrices appear in this case due to the structure of representations product, namely, in the tensor product $[r]\otimes [r]$ only one- and two-row Young diagrams appear. Correspondingly in the tensor cube of the representation one-, two- and three-row Young diagrams appear. Equation (\ref{e:3eig}) in fact allows one to find all the 3-strand Racah matrices from the known $U_q(sl_2)$ answer \cite{sl2} by setting $c=-n$.

In the context of this letter we discuss braids with larger number of strands. We will still only talk about symmetric representations, since for non-symmetric ones some of the $\mathcal{R}$-matrix eigenvalues start to coincide which makes eigenvalue conjecture unsolvable in the sense of (\ref{YBU}). Nevertheless known answers for Racah matrices show that eigenvalue conjecture still holds even in this case, but this case is out of the scope of this paper.

The claim of this paper is that from the 4-strand eigenvalue conjecture the following Racah matrices should coincide:
\setlength{\arraycolsep}{2pt}
\begin{equation}
\label{e:4eig}
%\begin{array}{l}
\left[
\begin{array}{ccc}
\ [r] & [p,q] & [a,b,d] \\
\ [r] & [k,l,m,n] & [x,y,z]
\end{array}\right]
=%\\
\end{equation}
%\\
\begin{equation*}
\left[
\begin{array}{ccc}
\ [r+c] & [p+c,q+c] & [a+c,b+c,d+c] \\
\ [r+c] & [k+c,l+c,m+c,n+c] & [x+c,y+c,z+c]
\end{array}\right],
%\end{array}
\end{equation*}

and similarly for larger number of strands:

\begin{equation}
\label{e:5eig}
%\begin{array}{l}
\left[
\begin{array}{ccc}
\ [r] & [p_1,\ldots p_i] & [a_1,\ldots a_{i+1}] \\
\ [r] & [k_1,\ldots k_{i+2}] & [x_1,\ldots x_{i+1}]
\end{array}\right]
=%\\
\end{equation}
%\\
\begin{equation*}
\left[
\begin{array}{ccc}
\ [r+c] & [p_1+c,\ldots p_i+c] & [a_1+c,\ldots a_{i+1}+c] \\
\ [r+c] & [k_1+c,\ldots k_{i+2}+c] & [x_1+c,\ldots x_{i+1}+c]
\end{array}\right].
%\end{array}
\end{equation*}

To explain these relations we need to check two things. First we will explain why the structure of tensor product of representations and, thusly, block structure of Racah matrices is the same. Then we will show that normalized eigenvalues are still the same.

\setlength{\arraycolsep}{5pt}

To understand why tensor product of representations remains the same, we prove the following property of tensor product of representations:
\begin{equation}
\begin{array}{c}
\ [a_1,\ldots, a_{n+1}] \in [p_1,\ldots,p_{n}]\otimes [r]
\\
\Updownarrow
\\
\ [a_1+1,\ldots, a_{n+1}+1] \in [p_1+1,\ldots,p_{n}+1]\otimes [r+1].
\end{array}
\label{EqRep}
\end{equation}
In this formula some of the elements of Young diagrams can be equal to zero. To prove this statement we have to explain which representations appear in the tensor product of representations. Basically \cite{Klimyk}, $[a_1,\ldots, a_{n+1}] \in [p_1,\ldots,p_{n}]\otimes [r]$ if and only if
\begin{equation}
\left\{
\begin{array}{l}
  a_i\leq p_i+r
  \\
  p_{i-1}\geq a_i \geq p_{i}
  \\
  a_1 \geq r
  \\
  \sum a_i=r+\sum p_i
\end{array}
\right.
\end{equation}
In fact the first line is superseded by the last one, since none of the elements $a_i$ can become smaller than $p_i$, while the whole size is changed by $r$. If we add or substract one from each element of Young diagrams then all of these properties but the first one hold identically, which shows that (\ref{EqRep}) indeed holds.

From (\ref{EqRep}) it follows that if we add one to each element of all the Young diagrams, then the whole tree of representations, like on Figs. \ref{f:3Rac}, \ref{f:4Rac1}, \ref{f:4Rac2}, remains the same, and thus also the block structure of the Racah matrices remains the same.

To explain the main claim of this paper, (\ref{e:4eig}) and (\ref{e:5eig}), we also need normalized eigenvalues of the $\mathcal{R}$-matrices to coincide. If one takes a product of two representations $[r]$ and representation $[2r-a,a]$ from this product then the corresponding eigenvalue is equal to
%coincides with representation $[2r-a+1,a+1]$ from the product of two representations $[r+1]$
\begin{equation}
\lambda_{[2r-a,a]}=q^{\varkappa_{[2r-a,a]}-4\varkappa_{[r]}-Nr}=q^{a^2-2ra}q^{r-Nr}.
\end{equation}
If we add one to each element of Young diagram, then it becomes
\begin{equation}
\begin{array}{r}
\lambda_{[2r-a+1,a+1]}=q^{\varkappa_{[2r-a+1,a+1]}-4\varkappa_{[r+1]}-Nr}=
\\
=q^{a^2-2ra}q^{-r+1-Nr-N}.
\end{array}
\end{equation}
Since we are interested in normalized eigenvalues, part independent of $a$ does not matter, while dependence on $a$ is the same. Thus the set of normalized eigenvalues is also the same.

Therefore we showed that the block structure of Racah matrices remain the same and the set of $\mathcal{R}$-matrix eigenvalues also remains the same if we add one to each element of all the Young diagrams. Thus the claims (\ref{e:4eig}) and (\ref{e:5eig}) also hold.

\section{Conclusion}

In this paper we discussed symmetries of Racah matrices and $6j$-symbols coming from the multi-strand eigenvalue conjecture. This allowed us to write down new symmetries of quantum $6j$-symbols, which are much harder to study using simpler 3-strand eigenvalue conjecture. We provide new symmetries involving non-symmetric representations. This is an important step on the way to understand the structure of quantum Racah matrices and, therefore, to understand representation theory of quantum groups. These Racah matrices then can also be applied to many other problems in theoretical in mathematical physics, especially in knot theory.

In this paper we discussed only braids with symmetric representations. However, the cases with non-symmetric representations are much more interesting. In this case there appear multiplicities in $\mathcal{R}$-matrices, which means that some eigenvalues start coinciding. This, of course, makes it much harder to understand the structure of Racah matrices, due to additional freedom in equation (\ref{YBU}). These matrices, however, seem to also possess block-structure, as was described in \cite{Mult}. This makes it look similar to the Racah matrices appearing in this paper. Such connection remain to be studied.

\section*{Acknowledgements}

We are grateful for very useful discussions to N. Tselousov and A. Sleptsov.

This work was supported by Russian Science Foundation grant No 20-12-00195.

\end{document}